\documentclass[12pt]{iopart}

\usepackage{graphicx}% Include figure files
\begin{document}

\title[Anisotropic flow of charged particles in ALICE]{Anisotropic flow of charged particles at $\mathbf{\sqrt{s_{NN}} = 2.76}$~TeV measured with the ALICE detector}
\author{Ante Bilandzic for the ALICE Collaboration}
\address{Nikhef,
Science Park 105,
1098 XG Amsterdam,
The Netherlands}
\address{Utrecht University, P.O. Box 80000, 3508 TA Utrecht, 
The Netherlands}
\ead{Ante.Bilandzic@nikhef.nl, abilandzic@gmail.com}

\begin{abstract}
Measurements of anisotropic flow in heavy-ion collisions provide evidence for the creation of strongly interacting matter which appears to behave as an almost ideal fluid. Anisotropic flow signals the presence of multiple interactions and is very sensitive to the initial spatial anisotropy of the overlap region in non-central heavy-ion collisions. In this article we report measurements of elliptic $v_2$, triangular $v_3$, quadrangular $v_4$ and pentagonal $v_5$ flow. These measurements have been performed with 2- and multi-particle correlation techniques. 
%We will show how these measurements can be understood from the initial spatial anisotropy and its fluctuations.
\end{abstract}

%Uncomment for PACS numbers title message
%\pacs{00.00, 20.00, 42.10}
% Keywords required only for MST, PB, PMB, PM, JOA, JOB? 
%\vspace{2pc}
%\noindent{\it Keywords}: Article preparation, IOP journals
% Uncomment for Submitted to journal title message
%\submitto{\JPA}
% Comment out if separate title page not required
%\maketitle

\vspace{-1 cm}

%======================================================================%
\section{Introduction}

Anisotropic flow \cite{Ollitrault:1992bk} measurements provide one of the most important tools to probe the properties of the medium generated in heavy-ion collisions. Anisotropic flow results at RHIC have yielded the evidence for the creation of strongly interacting matter, which appears to behave as an almost ideal fluid. In non-central heavy-ion collisions the initial volume of the interacting system is anisotropic in coordinate space. Due to multiple interactions this anisotropy is transferred to momentum space, and is then quantified via so-called flow harmonics $v_n$ \cite{Voloshin:1994mz}. In essence, anisotropic flow analysis is the measurement of flow harmonics $v_n$.
 
Due to the ellipsoidal collision geometry, the dominant harmonic in non-central collisions is $v_2$ ({\it elliptic flow}). In November 2010, 10 days after the first heavy-ion collisions were delivered by the LHC, the ALICE collaboration reported the initial measurement of $v_2$ at $\sqrt{s_{NN}} = 2.76$ TeV \cite{Aamodt:2010pa}. The primary new result was that integrated elliptic flow at LHC energies is about 30\% larger than at RHIC energies, while the differential elliptic flow $v_{2}(p_t)$ corresponds closely to the values measured at RHIC. The observed increase of $\sim\! 30\%$ in integrated elliptic flow is attributed to larger radial flow at LHC energies~\cite{Ref:Mikolaj}. In these proceedings we report on charged particle anisotropic flow measurements with an emphasis on the centrality dependence of $v_2$ and the new results for other harmonics.
% which have followed the published elliptic flow results \cite{Aamodt:2010pa}.

\vspace{-0.3 cm}

%======================================================================%
\section{Data sample and analysis}

The results presented in this article were obtained from a data sample comprising roughly 5M minimum bias Pb-Pb collisions at $\sqrt{s_{NN}} = 2.76$~TeV, after online and offline event selection was applied. Only the tracks within $|\eta|\! <\! 0.8$ and $0.2\!<\! p_{t}\! <\! 5.0$~GeV/$c$ and reconstructed with ALICE main tracking detector Time Projection Chamber (TPC) were used in the analysis. Tracks are rejected if their distance of closest approach to the primary vertex
is larger than 0.3 cm both in transverse and longitudinal direction. Finally, tracks are required to have at least 70 reconstructed space points in the TPC and a $\left<\chi^2\right>$ per TPC cluster $\leq 4$. Offline centrality determination utilized the VZERO detectors \cite{Ref:Alberica, Ref:Constantin}. 

Flow analysis with 2- and multi-particle azimuthal correlations has several well-known systematic biases, in particular the so-called {\it nonflow} contribution to azimuthal correlations. Nonflow is a systematic bias originating from the correlations involving only few particles. As first pointed out in \cite{Borghini:2001vi}, such nonflow correlations are largely suppressed by using multi-particle cumulants in the analysis. The other important systematic bias stems from multiplicity and flow fluctuations. The magnitude of multiplicity and trivial flow fluctuations, due to the size of centrality bins, can be estimated by comparing for instance the results of flow analysis in centrality bins of varying width. The systematic bias on $v_2$ due to multiplicity and trivial flow fluctuations was estimated and found to be negligible. Inefficiencies in the detector's azimuthal acceptance will strongly bias any measurement based on azimuthal correlations. Cumulants can be generalized to deal with such a systematic bias; however, a more practical approach used in this analysis was to use only the tracks reconstructed with TPC detector, making the bias negligible since the TPC has highly uniform azimuthal acceptance.

The analysis was performed using the cumulant method introduced by Ollitrault {\it et al} \cite{Borghini:2001vi}. This cumulant method was based on the formalism of generating functions, which has its own systematic biases and limitations. An improved version of cumulant method, which allows for fast and exact calculation of all multi-particle cumulants, was recently proposed in \cite{Bilandzic:2010jr}. To make a distinction between two versions of cumulant method the latter one is referred to as $Q$-cumulants (QC).

%======================================================================%
\section{Results}
\begin{figure}[h!]
\includegraphics[height=6.9cm,width=8cm]{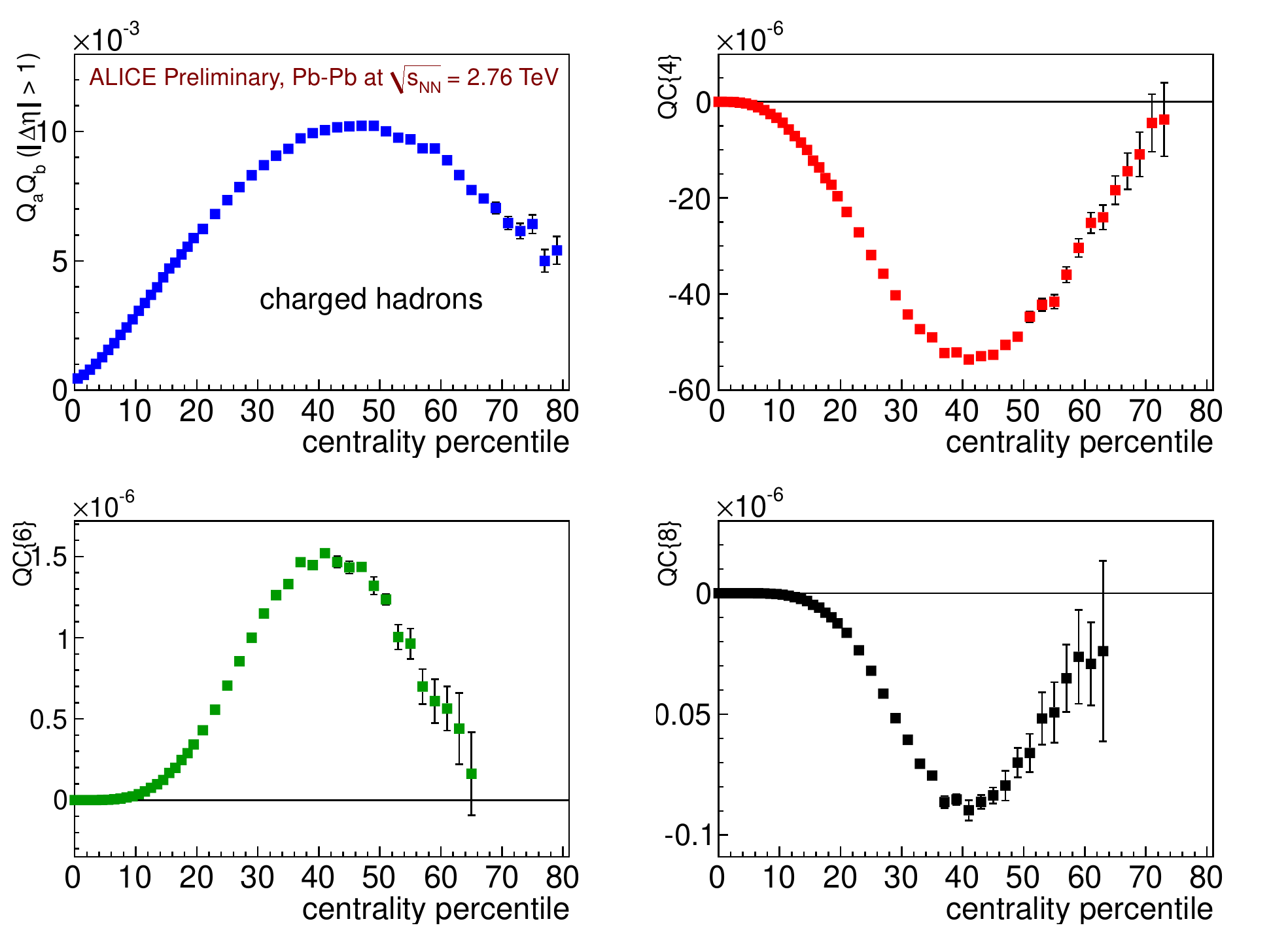}
\includegraphics[height=6.9cm,width=8cm]{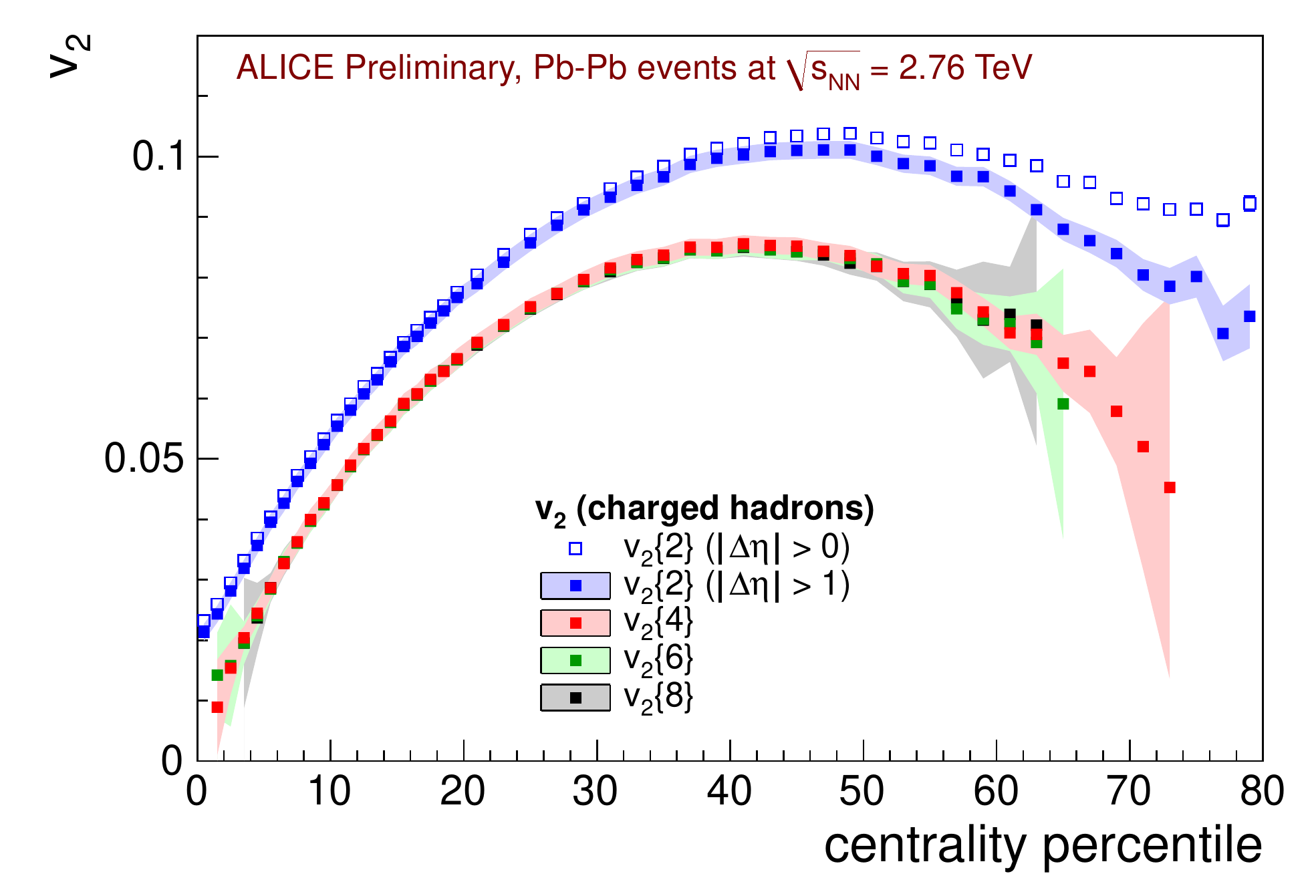}
\caption{The centrality dependence of cumulants measured for harmonic $n\!=\!2$ (left). The centrality dependence of $v_2$ estimated with different order cumulants (right).}
\label{fig:cumulants} 
\end{figure}

Flow contribution to cumulants is well understood and quantified \cite{Borghini:2001vi, Bilandzic:2010jr}:
%
%\begin{eqnarray}
%QC\{2\} &=& v_n^2\,,\nonumber\\
%QC\{4\} &=& -v_n^4\,,\nonumber\\
%QC\{6\} &=& 4v_n^6\,,\nonumber\\
%QC\{8\} &=& -33v_n^8\,.
%\label{QCs}
%\end{eqnarray}
%
%
\begin{eqnarray}
QC\{2\} = v^2\,, \qquad&QC\{4\} = -v^4 \,,\nonumber\\
QC\{6\} = 4v^6 \,,\qquad & QC\{8\} = -33v^8\,.
\label{QCs}
\end{eqnarray}
From above equations it is clear that in case the measured cumulants are dominated by flow correlations they will exhibit the characteristic flow signature $(+,-,+,-)$ irrespective of the order of harmonic for which cumulants are being measured. In Fig.~\ref{fig:cumulants} (left), cumulants for harmonic $n=2$ are shown for the centrality range 0-80\%, with centrality bin width of 1\% up to centrality 20\% and centrality bin width of 2\% for centralities beyond 20\% \cite{Ref:Alberica}. Clearly, the  cumulants exhibit the characteristic flow signature. Inversion of Eqs. (\ref{QCs})  yields an independent estimate for the harmonic $v_2$, to be denoted as $v_{2}\{2\}, v_{2}\{4\}, v_{2}\{6\}$ and $v_{2}\{8\}$, respectively. \footnote{To suppress nonflow contribution and to eliminate detector artifacts from reconstruction (e.g. track splitting), a $\left|\Delta\eta\right|$ gap was enforced among the particles being correlated in the measurement of the 2-particle cumulant.} These independent estimates for elliptic flow are presented in Fig.~\ref{fig:cumulants} (right). The 2-particle estimate, $v_{2}\{2\}$, was obtained by using two different  $\left|\Delta\eta\right|$ gaps (open and filled blue markers) in order to illustrate the magnitude of nonflow contribution. The nonflow contribution to the 2-particle cumulant scales as $\sim\! 1/(M\!-\!1)$, where $M$ is the multiplicity of the event, meaning that the relative nonflow contribution will be largest in the peripheral events where $M$ is smallest, as can be seen on Fig.~\ref{fig:cumulants} (right). On the other hand, all estimates from multi-particle cumulants are in an excellent agreement with each other (red, green and black markers) which indicates that already with the 4-particle cumulant nonflow is greatly suppressed so that there is little gain in suppressing it further by using 6- and 8-particle cumulants. This agreement between multi-particle cumulants is also due to the fact that, to leading order, they experience the same systematic bias due to statistical flow fluctuations. Finally, the difference between 2- and multi-particle estimates can be understood in terms of different (opposite in signature) sensitivity to the fluctuations in the initial geometry.

%________________________________________________________________________%
\begin{figure}[h!]
\includegraphics[height=6.9cm,width=8cm]{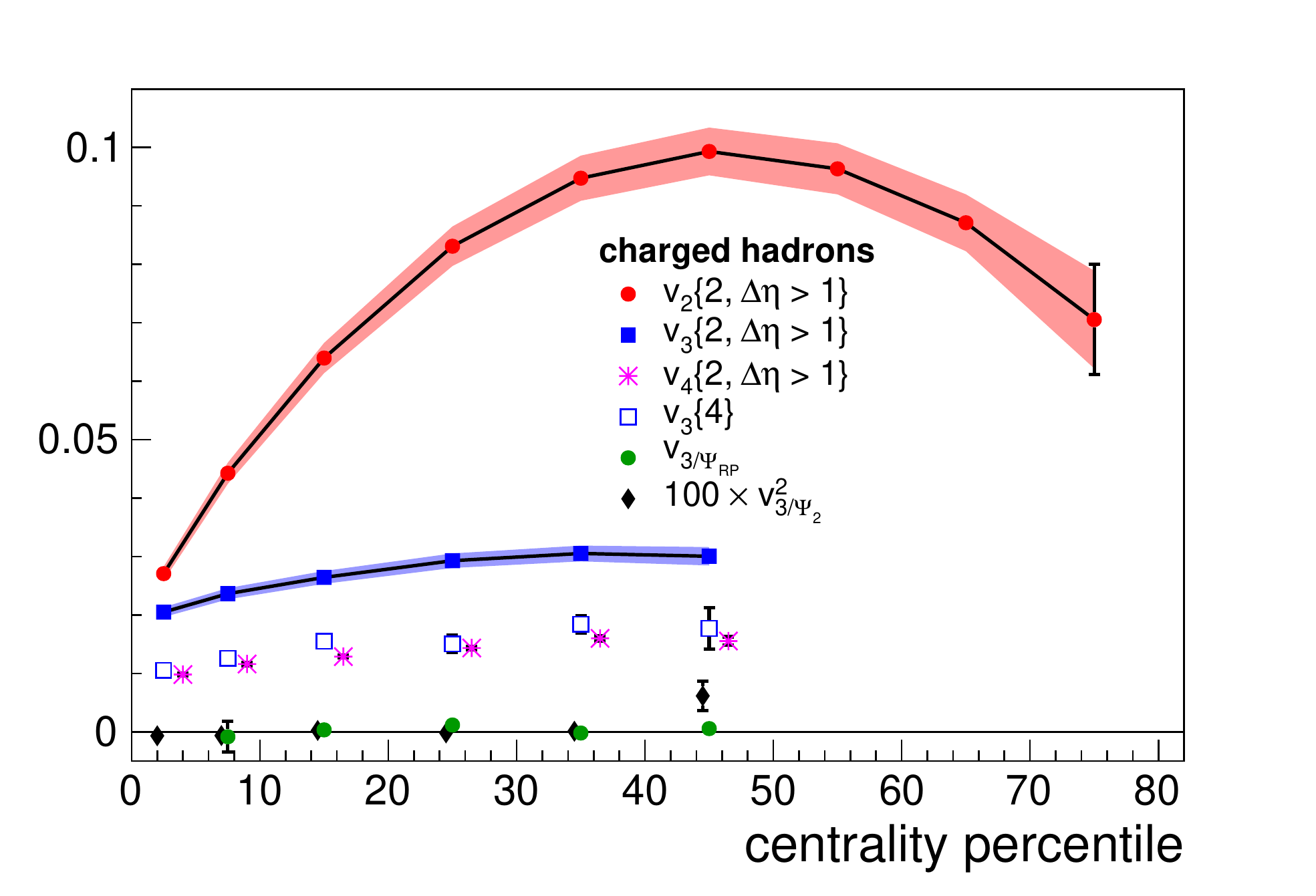}
\includegraphics[height=6.9cm,width=8cm]{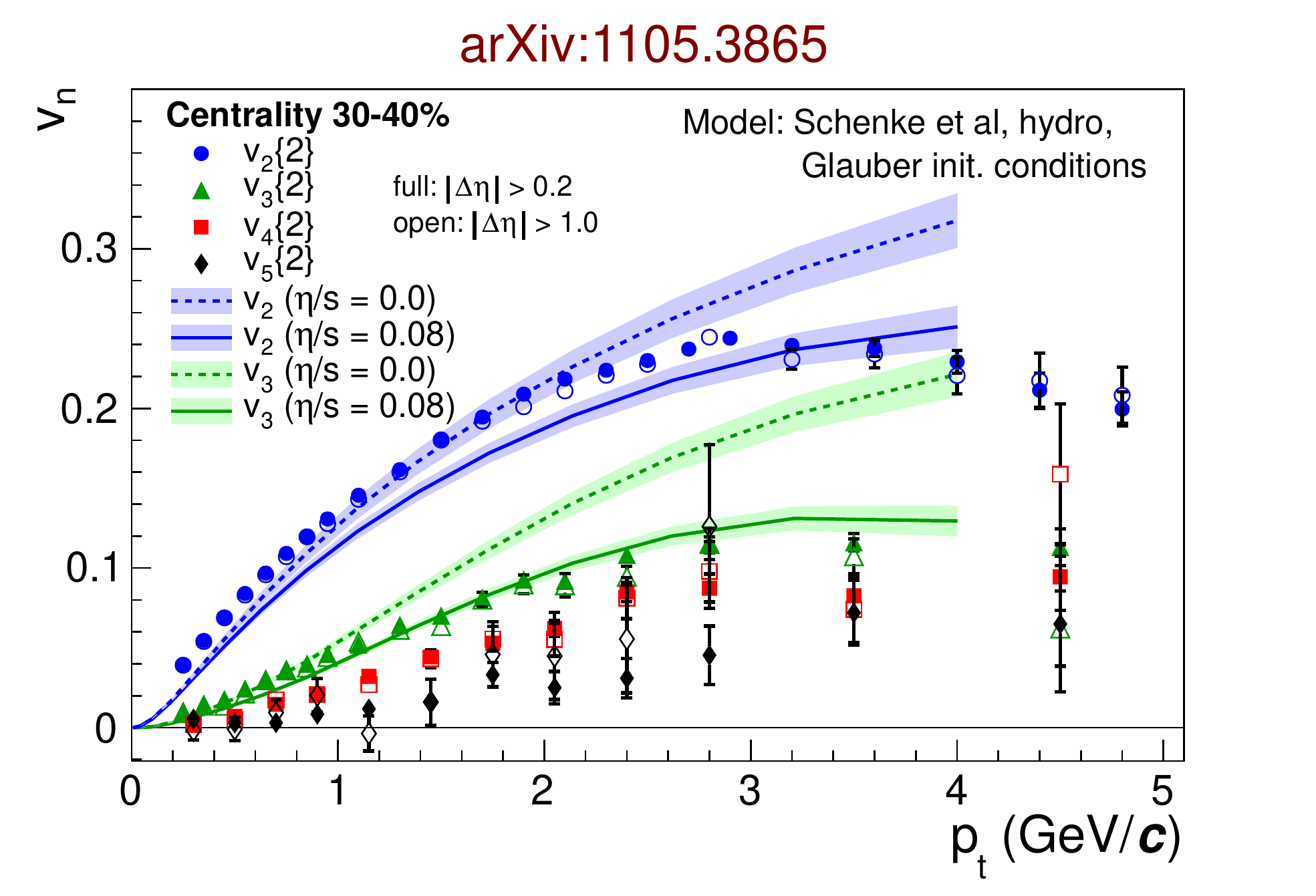}
\caption{Centrality (left) and $p_t$ dependence (right) of anisotropic flow harmonics~\cite{Collaboration:2011vk}.}
\label{fig:fig1a_fig3a} 
\end{figure}
%________________________________________________________________________%

In Fig.~\ref{fig:fig1a_fig3a} (left) the new results for harmonics beyond $v_2$ are presented~\cite{Collaboration:2011vk}. The primary new result is that triangular flow, $v_3$, is not zero. $v_3$ cannot develop as a correlation of all particles with the reaction plane (the event-ensemble average of odd harmonics must be zero at mid-rapidity due to the symmetry of collision), but instead as a correlation with the participant plane of $v_3$ (event-by-event fluctuations in the initial geometry yields event-by-event triangularity in coordinate space which  determines such symmetry plane). A new picture of flow analysis has recently emerged according to which each harmonic has its own participant plane, raising the question of the relation between the different participant planes. As an example, in Fig.~\ref{fig:fig1a_fig3a} (left) we demonstrate that the participant planes of $v_2$ and $v_3$ are uncorrelated (black diamonds). Triangular flow estimated with 4-particle cumulant (open blue squares) is half as large as the corresponding 2-particle estimate (filled blue squares), in agreement with a recent prediction by Bhalerao {\it et al} \cite{Bhalerao:2011yg}.

In Fig.~\ref{fig:fig1a_fig3a} (right), the results for $p_t$ dependence of harmonics $v_2, v_3, v_4$ and $v_5$ are presented~\cite{Collaboration:2011vk}. For  elliptic and triangular flow, comparison is made with a models based both on ideal and viscous relativistic hydrodynamics, respectively, with Glauber initial conditions (taking into account the role of event-by-event fluctuations of the initial conditions) as was proposed in~\cite{Schenke:2011tv}. Within this particular model the overall magnitude of elliptic and triangular flow agrees with the measurement, but the details of the $p_t$ dependence are not well described. In particular, the magnitude of $v_2(p_t)$ is described better with ideal hydro with $\eta /s = 0$, while for $v_3(p_t)$ the model with $\eta /s = 0.08$ provides a better description, meaning that this model fails to describe well $v_2$ and $v_3$ simultaneously. 

\vspace{-0.3 cm}
%======================================================================%
%\section{Summary}
%
%The new results of anisotropic flow analysis for Pb-Pb collisions at LHC in ALICE were presented and %comparison to theoretical predictions has been made.


\vspace{-0.3 cm}
\section*{References}
\begin{thebibliography}{99}
\bibitem{Ollitrault:1992bk}
 J.~Y.~Ollitrault,
 %``Anisotropy As A Signature Of Transverse Collective Flow,''
 Phys.\ Rev.\  D {\bf 46} (1992) 229.
 %%CITATION = PHRVA,D46,229;%%

%\cite{Voloshin:1994mz}
\bibitem{Voloshin:1994mz}
  S.~Voloshin and Y.~Zhang,
  %``Flow study in relativistic nuclear collisions by Fourier expansion of
  %Azimuthal particle distributions,''
  Z.\ Phys.\  C {\bf 70} (1996) 665.
  %[arXiv:hep-ph/9407282].
%[arXiv:\href{http://arxiv.org/abs/hep-ph/9407282}{hep-ph/9407282}].
  %%CITATION = ZEPYA,C70,665;%%

%\cite{Aamodt:2010pa}
\bibitem{Aamodt:2010pa}
  K.~Aamodt {\it et al.}  [The ALICE Collaboration],
  %``Elliptic flow of charged particles in Pb-Pb collisions at 2.76 TeV,''
  Phys.\ Rev.\ Lett.\  {\bf 105}, 252302 (2010).
  %arXiv:1011.3914 [nucl-ex].
  %%CITATION = ARXIV:1011.3914;%%

\bibitem{Ref:Mikolaj} {M.~Krzewicki, these proceedings.}

\bibitem{Ref:Alberica} {A.~Toia, these proceedings.}

\bibitem{Ref:Constantin} {C.~Loizides, these proceedings.}

%\cite{Borghini:2001vi}
\bibitem{Borghini:2001vi}
  N.~Borghini, P.~M.~Dinh and J.~Y.~Ollitrault,
  %``Flow analysis from multiparticle azimuthal correlations,''
  Phys.\ Rev.\  C {\bf 64} (2001) 054901.
  %[arXiv:nucl-th/0105040].
%[arXiv:\href{http://arxiv.org/abs/nucl-th/0105040}{nucl-th/0105040}].
  %%CITATION = PHRVA,C64,054901;%% 

%\cite{Bilandzic:2010jr}
\bibitem{Bilandzic:2010jr}
  A.~Bilandzic, R.~Snellings, S.~Voloshin,
  %``Flow analysis with cumulants: Direct calculations,''
  Phys.\ Rev.\  C {\bf 83} (2011) 044913.
  %[arXiv:1010.0233 [nucl-ex]].

%\cite{Collaboration:2011vk}
\bibitem{Collaboration:2011vk}
  K.~Aamodt {\it et al.}  [The ALICE Collaboration],
  %``Higher harmonic anisotropic flow measurements of charged particles in Pb-Pb collisions at 2.76 TeV,''
 [arXiv:1105.3865 [nucl-ex]].

 %\cite{Bhalerao:2011yg}
\bibitem{Bhalerao:2011yg}
  R.~S.~Bhalerao, M.~Luzum and J.~Y.~Ollitrault,
  %``Determining initial-state fluctuations from flow measurements in heavy-ion
  %collisions,''
  [arXiv:1104.4740 [nucl-th]].
  %%CITATION = ARXIV:1104.4740;%%

  %\cite{Schenke:2011tv}
\bibitem{Schenke:2011tv}
  B.~Schenke, S.~Jeon, C.~Gale,
  %``Anisotropic flow in sqrt(s)=2.76 TeV Pb+Pb collisions at the LHC,''
  [arXiv:1102.0575 [hep-ph]].

\end{thebibliography}
\end{document}